\documentclass[aps,prl,reprint,superscriptaddress]{revtex4-1}
\usepackage[utf8]{inputenc}
\usepackage[dvips]{graphicx}
\usepackage{bm}
\usepackage{amsmath}
\usepackage{xcolor}
\usepackage{hyperref}

\begin{document}

\title{Detecting parity effect in a superconducting device in the presence of parity switches 
}

 \author{E. T. Mannila}
 \email{elsa.mannila@aalto.fi}
 \affiliation{QTF Centre of Excellence, Department of Applied Physics, Aalto University, FI-00076 Aalto, Finland}
 \author{V. F. Maisi}
 \affiliation{QTF Centre of Excellence, Department of Applied Physics, Aalto University, FI-00076 Aalto, Finland}
 \affiliation{Division of Solid State Physics and NanoLund, Lund University, 22100 Lund, Sweden}
 \author{H. Q. Nguyen}
 \affiliation{Center for Quantum Devices and Station Q Copenhagen, Niels Bohr Institute, University of Copenhagen, Copenhagen, Denmark}
 \affiliation{Nano and Energy Center, Hanoi University of Science, VNU, 120401 Hanoi, Vietnam}
 \author{C. M. Marcus}
 \affiliation{Center for Quantum Devices and Station Q Copenhagen, Niels Bohr Institute, University of Copenhagen, Copenhagen, Denmark}
 \author{J. P. Pekola}
 \affiliation{QTF Centre of Excellence, Department of Applied Physics, Aalto University, FI-00076 Aalto, Finland}

\date{\today}
\begin{abstract}
We present a superconducting device showing a clear parity effect in the number of electrons, even when there is, on average, a single nonequilibrium quasiparticle present and the parity of the island switches due to quasiparticles tunneling in and out of the device at rates on the order of 100 Hz. We detect the switching by monitoring in real time the charge state of a superconducting island connected to normal leads by tunnel junctions. The quasiparticles are created by Cooper pairs breaking on the island at a rate of a few kHz. We demonstrate that the pair breaking is caused by the backaction of the single-electron transistor used as a charge detector. With sufficiently low probing currents, our superconducting island is free of quasiparticles 97\% of the time. 
\end{abstract}

\maketitle

In a superconductor, electrons participating to conduction form Cooper pairs. The minimum energy for an unpaired quasiparticle excitation is $\Delta$, the superconducting gap, which leads to a free energy difference between states with an even and odd number of electrons in the absence of subgap states. The resulting parity effect is commonly observed in features periodic in $2e$, with $e$ the electron charge, in the transport through an island or by measuring the average charge in an isolated box \cite{tuominen1992, eiles1993, lafarge1993measurement, lafarge1993two}. In thermal equilibrium, the parity effect disappears at temperatures where quasiparticles are excited, around 200 mK for typical micron-scale aluminum structures, as the free energy difference disappears. A clean $2e$ periodicity of Coulomb blockade is often taken to suggest a device free of quasiparticles \cite{macleod2009, cedergren2015, vaitiekenas2018}. 

In addition to suppressing the parity effect \cite{aumentado2004, mannik2004}, quasiparticle excitations are generally detrimental for superconducting devices.
In Josephson junction based qubits, quasiparticles tunneling across the junction cause decoherence \cite{martinis2009, catelani2011relaxationprl}. For quantum computing using Majorana modes in superconductor-semiconductor hybrids, topological protection is only present when the total fermion parity of the system stays constant. The parity lifetime is a fundamental bound to the coherence time of such a qubit \cite{rainis2012, higginbotham2015, albrecht2017}. 
At low temperatures the quasiparticle density $n_{qp}$ should be exponentially suppressed, and the parity lifetime consequently exponentially long. In practice, often a saturation of $n_{qp}$ to values several orders of magnitude higher than in thermal equilibrium is observed  in experiments on qubits \cite{shaw2008, martinis2009, riste2013}, resonators \cite{devisser2011, barends2011, grunhaupt2018}, and quantum capacitance \cite{stone2012} and kinetic inductance detectors \cite{day2003}. Another quantity related to $n_{qp}$ is the poisoning time between successive quasiparticle tunneling events. Quasiparticle densities or poisoning times can be inferred from transport measurements \cite{knowles2012, maisi2013, vanwoerkom2015, higginbotham2015, albrecht2017} or qubit coherence times \cite{lenander2011}. Even-to-odd transitions from quasiparticle tunneling can also be measured in real time with radio-frequency reflectometry \cite{naaman2006, ferguson2006, court2008, shaw2008} or in the parity-dependent frequency shift of transmon qubits \cite{riste2013, serniak2018}. In this work, we measure in real time quasiparticle tunneling and parity switching on a superconducting island, and observe a parity effect in the parity-dependent occupation probabilities and tunneling rates of charge states. The quasiparticles are created from the backaction of the charge detector \cite{mannik2004}. This is a critical issue for Majorana qubit proposals incorporating charge readout \cite{aasen2016, karzig2017}. 

\begin{figure}
\includegraphics{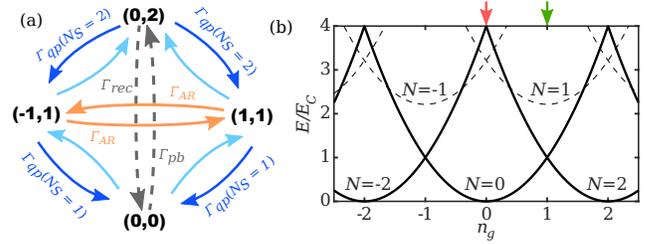}
\caption{
 (a) States $(N,N_S)$ with $N$ excess charges and $N_S$ quasiparticle excitations on the superconducting island. Quasiparticles tunneling from the island to normal metal leads at rates $\Gamma_{qp}(N_S)$ given by Eq. \eqref{eq:gqp} are directly detected, while quasiparticle tunneling into the island (light blue) is suppressed by the superconducting gap. 
Cooper pairs break at a rate $\Gamma_{pb}$ and recombine with $\Gamma_{rec}$. Andreev tunneling events ($\Gamma_{AR}$) transfer two electrons on or off the island while keeping the number of quasiparticles constant. 
 (b) The parity-dependent free energy $E = E_C(N-n_g)^2 + F(T_S) \times N \bmod 2$ of even (solid lines) and odd (dashed) charge states $N$, calculated at $k_B T_S/\Delta= 0.02$ and $E_C/\Delta = 0.33$. Arrows show values of the gate offset $n_g$ for the charge detector traces shown in Fig. \ref{sampletraces}.
 \label{physics}}
\end{figure}

We characterize the state of the superconducting island with the excess charge $N$ and number of excitations $N_S$ on the island, where $N$ and $N_S$ are integers of the same parity, following Ref. \cite{maisi2013}, which provides the quantitative details of the model. The relevant processes in our system are shown in Fig. \ref{physics}(a).  If there are two or more excitations on the island, they can recombine to a Cooper pair with rates $\Gamma_{rec}(N_S)$. We include recombination via the electron-phonon coupling. Cooper pairs on the island can break, creating two excitations, with a rate $\Gamma_{pb}$, assumed independent of the state of the island. 
The thermal electron-phonon pairbreaking rate is vanishingly small at the temperatures of the experiment, so this rate arises from nonequilibrium conditions.
We directly detect the quasiparticle tunneling events between the superconducting island and normal metal leads at temperature $T_N$, which change both $N$ and $N_S$ by one. If excess quasiparticles are present (the superconductor temperature $T_S > T_N$) but  $k_B T_N \ll \Delta$, the rate for quasiparticles tunneling out of the island $\Gamma_{qp}(N_S) \equiv \Gamma(N \rightarrow N \pm 1, N_S \rightarrow N_S -1)$ depends, for a range of energy gains, only on the quasiparticle density $n_{qp} = \sqrt{2 \pi} D(E_F) \sqrt{\Delta k_B T_S} e^{-\Delta/k_B T_S}$ or $N_S = n_{qp} V$ before the tunneling event as \cite{saira2012}
\begin{equation}\label{eq:gqp}
\Gamma_{qp}(N_S) = \frac{N_S}{2 e^2 R_T D(E_F) V}.
\end{equation}
Here, $R_T$ is the resistance of the tunnel junction, $k_B$ the Boltzmann constant, and $D(E_F) = 2.15 \times 10^{47}$ J$^{-1}$ m$^{-3}$ \cite{devisser2011} the normal density of states (including spin) at the Fermi level. 
In particular, a single quasiparticle in the island with volume $V= 550$ nm $\times$ 2 $\mu$m $\times$ 50 nm and $R_T = 15.6$ M$\Omega$ corresponds to $\Gamma_{qp}$ = 110 Hz.
Tunneling events which increase $N_S$ are suppressed by the superconducting gap when $N$ is close to the gate offset $n_g$. 

If $\Gamma_{pb}$ is zero in the model above, we recover the thermal equilibrium case. The free energies of the charge states $N$ of a superconducting island are $E = E_C(N-n_g)^2 + F(T_S) \times N \bmod 2$, which includes, in addition to the contribution of the charging energy $E_C=e^2/2 C_\Sigma$ with $C_\Sigma$ the total capacitance of the island, the free energy cost $F(T_S) \approx \Delta -k_B T_S \ln (D(E_F) V \Delta)$ of an unpaired excitation \cite{tuominen1992, lafarge1993measurement}. The approximation is valid when $k_B T_S \ll \Delta$. These free energies are sketched in Fig. \ref{physics}(b) against $n_g = C_g V_g/e$, where $V_g$ is the voltage applied to a gate electrode coupled via capacitance $C_g$. When $F(T_S) > E_C$, as in our devices below 120 mK, the ground state has even parity, and we expect to see only two-electron Andreev tunneling events. 
The states with odd $N$ should become significantly occupied only above the temperature $T_0 = \Delta / (k_B \ln(V D(E_F) \Delta)) \approx $ 190 mK, where a single quasiparticle is thermally excited on the island. 

\begin{figure}
\includegraphics{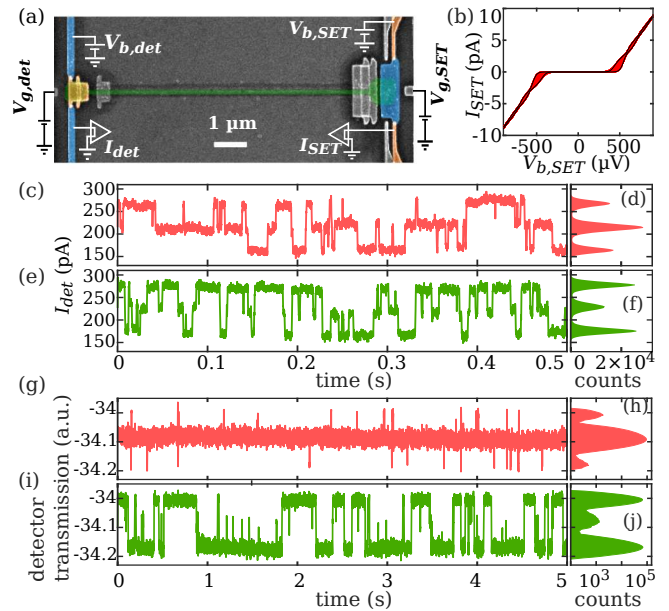}
\caption{ (a) False color scanning electron micrograph of the device and measurement diagram for sample A. The superconducting aluminum island (blue) connected to normal copper leads (orange) is shown on the right. The charge detector on the left side of the image is coupled to the island with a chromium wire (green) under a 40 nm insulating aluminum oxide layer grown by atomic layer deposition. The devices are biased with voltages $V_{b,SET/det}$, while the current through the device is recorded with a room-temperature current amplifier. Gate voltages $V_{g,SET/det}$ tuning the gate offsets are applied to capacitively coupled electrodes.
 (b) Large-scale current-voltage characteristics of the superconducting island of sample A, measured (red) over several periods in the gate voltage $V_{g,SET}$ for each bias voltage $V_{b,SET}$. Black lines are simulations at $n_g=0$ and $n_g=0.5$ with parameters given in the text. 
(c,e) Real-time detector output of sample A at $V_{b,SET}=0$ and $V_{b,det}=490$ $\mu$eV with (c) $n_g=0$ and (e) $n_g = 1$. Three charge states are occupied in both cases, but even charge states have a higher occupation probability.
(d,f) Histograms of the traces in panels (c,e).
(g,i) Real-time detector output of sample B at $V_{b,SET}=0$ and $V_{b,det}=350$ $\mu$eV. (h,j) Histograms of the traces in panels (g,i). Note that in contrast to (d,f), a logarithmic scale is used to show the minuscule occupation of the odd states.
\label{sampletraces}
}
\end{figure}

Our device, shown in Fig. \ref{sampletraces}(a), is a single-electron transistor (SET) with a superconducting aluminum island connected to normal metal copper leads with aluminum oxide tunnel barriers a few nanometers thick.  
The capacitively coupled charge detector is another SET, but with a copper island and aluminum leads. The devices were fabricated with standard electron-beam lithography and three-angle evaporation on thermally oxidized silicon substrates. We have measured two similar devices, samples A and B. 
The energy gap $\Delta=206$ $\mu$eV (210 $\mu$eV), total tunnel resistance of the two junctions $70$ M$\Omega$ (40 M$\Omega$) and the charging energy $E_C = 0.33\Delta=68$ $\mu$eV ($0.45\Delta =$ 95 $\mu$eV) of sample A (B) were determined by fitting the current-voltage characteristics 
as shown in Fig. \ref{sampletraces}(b). For the electron counting experiments, the superconducting island was kept at zero bias. The island acts as a single-electron box connected to normal leads through the parallel resistance of the two junctions $R_T = 15.6$ M$\Omega$ (8.9 M$\Omega$), with both devices having unequal tunnel junctions whose areas and resistances differ by a factor of 2. 
Sample A was measured 
at 60 mK in a DC measurement setup sketched in Fig. \ref{sampletraces}(a), where we directly record the amplified detector current $I_{det}$. Sample B was measured 
at 25 mK in a setup where the detector was used as an RF-SET \cite{schoelkopf1998} \footnote{See Supplemental Material, which contains Refs. \cite{zorin1995, gasparinetti2015, viisanen2015, pekola2010}, for details on the measurement setup, filtering and shielding, and fits for Sample B.}. \nocite{zorin1995, gasparinetti2015, viisanen2015, pekola2010}

Figure \ref{sampletraces}(c-j) shows real-time traces of the charge detector output at $n_g=0$ (c,d,g,h) and $n_g=1$ (e,f,i,j). In sample A (Fig. \ref{sampletraces}(c-f)), three charge states are always occupied for a significant fraction of time, even though the charging energy ($E_C/k_B \approx $ 800 mK) is much larger than the bath temperature. Most of the transitions are single-electron transitions. 
At $n_g=0$ the state at $I_{det} = $ 200 pA corresponding to $N=0$ is more occupied than $N=\pm 1$ at 150 pA and 250 pA, while at $n_g=1$ the state $N=1$ (220 pA) has a lower occupation probability than $N=0$ (170 pA) or $N=2$ (270 pA). In sample B (Fig. \ref{sampletraces}(g-j)), where using smaller detector currents is possible (see supplemental material \cite{Note1}, the odd states are occupied with almost two orders of magnitude lower probability, suggesting a much lower density of nonequilibrium quasiparticles. 

\begin{figure}
\includegraphics{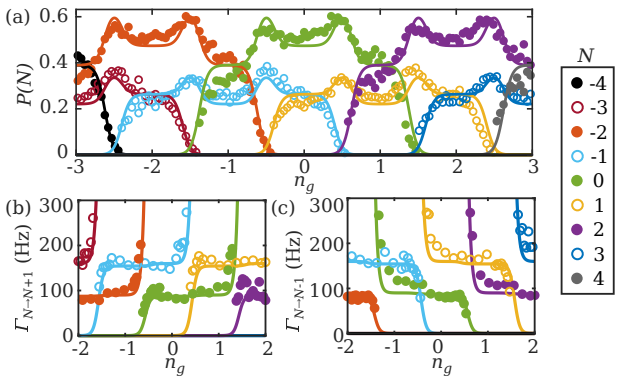}
\caption{
(a) Occupation probabilities $P(N)$  and (b,c) tunneling rates  $\Gamma_{N,N\pm 1}$ between charge states $N$ over a range of $n_g$ in sample A. Filled (open) circles are measured data for even (odd) $N$. There are always at least three charge states occupied with more than 10\% probability, a nonequilibrium situation, but the most probable state always has even parity. 
The transition rates have plateaus at $\Gamma_{even}=80$ Hz and $\Gamma_{odd}=160$ Hz with the rate depending only on the parity of the initial state. This corresponds to a parity-dependent quasiparticle density on the island.
Solid lines are simulations with the Cooper pair breaking rate $\Gamma_{pb} = 4.6$ kHz as the only free parameter. 
\label{analysis}
}
\end{figure}

We measure time traces across a range of $n_g$ and extract the occupation probabilities of each charge state (Fig. \ref{analysis}(a)). At all values of $n_g$ there are three or four charge states visible. This can be explained with a nonequilibrium quasiparticle population. Intuitively, if there is a quasiparticle with energy $\Delta=3E_C$ on the island, the energy cost of charging the island with an additional electron is possible to overcome. If $n_g=0.5$, the charging part of the energy $E_C(N-n_g)^2$ is smaller than $\Delta$ for the states $N=-1, 0, 1$ and 2, which are the states observed. However, even in this nonequilibrium situation, the most probable state has always even parity as in thermal equilibrium.

The tunneling rates $\Gamma_{N\rightarrow M}$ of single-electron transitions (Fig. \ref{analysis}(b-c)) are determined from the time traces. 
For each transition, there is a range in $n_g$ where the rate is independent of the energy gained in the transition, since the tunneling rates are dominated by excess quasiparticles in the superconductor 
 \cite{saira2012}.
The measured rates $\Gamma_{N\rightarrow N \pm 1}$ are a weighted average of the rates $\Gamma_{qp} (N_S)$ over the $N_S$ states for a given $N$ 
and thus directly proportional to a mean quasiparticle number $\langle N_S \rangle$. 
At the plateaus, $\Gamma_{qp}( \langle N_{S,odd} \rangle ) = $ 160 Hz for odd $N$ and $\Gamma_{qp}( \langle N_{S,even} \rangle ) = $ 80 Hz for even $N$, and thus the mean quasiparticle population depends on parity. The ratio $\Gamma_{qp}( \langle N_{S,odd} \rangle )/ \Gamma_{qp}( \langle N_{S,even} \rangle ) = \langle N_{S,odd} \rangle / \langle N_{S,even} \rangle \approx 2$ means that $\langle N_{S,even} \rangle \geq 0.5$ and at least two quasiparticles must be present for 25\% of the time in even charge states. 
To maintain such a quasiparticle population, quasiparticles must be generated either from Cooper pairs breaking or electrons tunneling from the leads with a total rate on the same order as with what they tunnel out or recombine. The expected recombination rate is 9.7 kHz for $N_S = 2$ and larger for more quasiparticles (assuming the electron-phonon coupling constant $\Sigma = 1.8 \times 10^9$ W K$^{-5}$m$^{-3}$ \cite{maisi2013}), two orders of magnitude larger than the measured tunneling rates. A Cooper pair breaking rate much larger than the tunneling rates is then needed to produce the observed excess quasiparticles, in contrast to models where quasiparticles tunnel in from the leads \cite{aumentado2004, vanveen2018}. Any broken Cooper pair will, on average, recombine on the superconducting island before having time to tunnel out. The quasiparticle population on the island is determined by the competition between pair breaking and recombination, with the tunnel contacts only serving to probe the resulting quasiparticle density.

We calculate numerically the transition rates between different $(N, N_S)$ states as in Ref. \cite{maisi2013}, which gives the quasiparticle tunneling and recombination rates and a corresponding rate equation, and solve for the steady-state occupation probabilities. The solid lines in Fig. \ref{analysis}(a) are the occupation probabilities for each charge state with any number of excitations, while the solid lines in Figs. \ref{analysis}(b-c) are the average rates between different charge states. To reproduce the significant occupation probabilities of odd charge states, we need to include a Cooper pair breaking rate $\Gamma_{pb} = 4.6 $ kHz. 
We are able to reproduce quantitatively all the features in the transition rates and occupation probabilities with only $\Gamma_{pb}$ as a free parameter in our model.
Other parameters are either determined from independent measurements ($R_T$, $\Delta$, $E_C$) or they are known literature values ($\Sigma$, $D(E_F)$). 
Some of the transitions interpreted as two successive single-electron events might be two-electron Andreev events, 
which are not included in the model. Yet their influence to obtained results is weak: assuming successive transitions from $N$ to $N \pm 2$ occurring within 1 ms to be Andreev events decreases the inferred single-electron tunnel rates only by a few percent.
The finite bandwidth of the detector (a few kHz) mostly causes the measured rates to underestimate the true rates at the quasiparticle-induced plateaus by 10\% to 20\% \cite{naaman2006poisson} and does not affect our main conclusions. In Fig. \ref{analysis}, the simulated rates are corrected to account for finite bandwidth using the model of Ref. \cite{naaman2006poisson}. 

The time-averaged number of excitations from the simulations is $\langle N_S \rangle = 0.86$ in even charge states and 1.6 in odd states. The equilibrium temperature where the parity effect is expected to disappear corresponds to a single quasiparticle being excited. It is somewhat against the common view that the parity effect is clearly visible even with a single nonequilibrium excitation present and quasiparticles continuously tunneling in and out of the device, as we demonstrate here. 
The ratio of the pair breaking and recombination rates $\Gamma_{pb}/\Gamma_{rec}$ determines the quasiparticle density. However, we cannot determine these two rates independently: the agreement between simulations and experiment in Fig. \ref{analysis} remains equally good if $\Gamma_{pb}$  and the electron-phonon coupling constant $\Sigma$ setting $\Gamma_{rec}$ are scaled up or down but by the same factor. 

\begin{figure}
\includegraphics{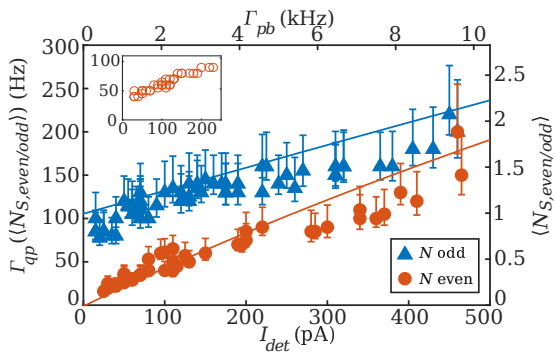}
\caption{ Quasiparticle tunneling rates $\Gamma_{qp}$ in sample A corresponding to events out of charge states with even ($\Gamma_{qp}(\langle N_{S,even} \rangle )$, circles) and odd ($\Gamma_{qp}(\langle N_{S,odd} \rangle )$, triangles) parity. $\Gamma_{qp}$ corresponds to a mean quasiparticle number $\langle N_S \rangle$ according to Eq. \eqref{eq:gqp}. The rates decrease with decreasing detector current, and $\Gamma_{qp}(\langle N_{S,even} \rangle )$ extrapolates to less than 15 Hz with $I_{det} = 0$. 
Solid lines are  calculated $\langle N_S \rangle$ for $N=0$ (red) and $N=1$ (blue) at $n_g=0.5$ as a function of the Cooper pair breaking rate $\Gamma_{pb} = A I_{det}/e$. 
Inset: $\Gamma_{qp}(\langle N_{S,even} \rangle )$ measured two weeks earlier extrapolates to 30 Hz at $I_{det} = 0$, which corresponds to $\Gamma_{pb} \approx$ 1 kHz. 
    \label{backaction}
}
\end{figure}

We now turn to the origin of the Cooper pair breaking rate observed. We repeat the measurement of time traces versus $n_g$ in sample A at different $V_{b,det}$ between 385 $\mu$V and 560 $\mu$V and extract $\Gamma_{qp}(\langle N_{S,even} \rangle )$ and $\Gamma_{qp}(\langle N_{S,odd} \rangle )$ at the quasiparticle-induced plateaus, shown as a function of $I_{det}$ in Fig. \ref{backaction}. The tunneling rates increase linearly with detector current, but $\Gamma_{qp}(\langle N_{S,even} \rangle )$ extrapolates to $10\pm 7$ Hz at $I_{det}=0$ and $\Gamma_{qp}(\langle N_{S,odd} \rangle )$ to somewhat below 100 Hz, close to the calculated tunneling rate 110 Hz of one quasiparticle. 
To model pair breaking by backaction, we assume $\Gamma_{pb} = A I_{det}/e$ without any detector-independent rate. The fit parameter $A = 1/ 300,000$ is the probability for an electron tunneling in the detector to break a Cooper pair. 
We calculate the mean quasiparticle number in even and odd states at $n_g=0.5$ as a function of $\Gamma_{pb}$ (Fig. \ref{backaction}) and convert it to a tunneling rate using Eq. \eqref{eq:gqp}. 
Fitting the occupation probabilities and tunneling rates as in Fig. \ref{analysis} to measurements at different $V_{b,det}$ confirms that the effect of the detector is only to break Cooper pairs, as no other parameters need to be changed for a good fit (data not shown). A similar linear dependence on the current of a nearby SET was observed as quasiparticle poisoning rate of a fully superconducting SET in Ref. \cite{mannik2004}. Nonequilibrium phonons created in the superconducting parts of the detectors 
\cite{otelaja2013} could be a plausible mechanism to explain the observed dependence on $I_{det}$ \cite{patel2017}, but we cannot rule out a photon-mediated backaction mechanism \cite{naaman2007, lotkhov2012}. We have also observed a reduction of $\Gamma_{qp}(\langle N_{S,even} \rangle )$ extrapolated at $I_{det} = 0$  from 30 Hz to 10 Hz during a 3-week cooldown due to unknown reasons (Fig. \ref{backaction} inset). 

In sample B, the mean quasiparticle numbers $\langle N_S \rangle$ decrease with decreasing $V_{b,det}$, qualitatively similarly as in sample A. By fitting data similar to that shown in Fig. \ref{analysis} \cite{Note1}, we obtain $\Gamma_{pb}$ = 100 Hz at $V_{b,det} = 350$ $\mu$eV, which corresponds to $\langle N_S \rangle$ = 0.04, $n_{qp}= \langle N_S \rangle/V$ = 0.7 $\mu$m$^{-3}$ and zero excitations on the island for 97\% of the time. The zero-bias conductance of the device would appear 2$e$ periodic although the parity switches several times a second. Measuring $I_{det}$ directly is not possible in this setup, but the detector of sample B has higher sensitivity at small $V_{b,det}$, allowing smaller detector currents (estimated $I_{det}<$ 10 pA). Therefore, we attribute the decreased $\langle N_S \rangle$ in sample B to reduced detector backaction compared to sample A. 

In conclusion, we have observed a clear parity effect in the occupation probabilities and tunneling rates of the charge states of a superconducting island, even in the presence of a single nonequilibrium excitation and frequent parity switches. The excitations are generated by Cooper pairs breaking on the superconducting island, and the quasiparticles almost always recombine before tunneling out.  The poisoning time or parity lifetime of the island---defined as the time between quasiparticle tunneling events---can be long, even though the island is still poisoned in the sense of quasiparticles being present.  The Cooper pair breaking is caused by the backaction of the charge detector, which can be minimized by reducing the detector current. 
We expect that in future experiments, the statistics of electron counting yields access to the recombination and pair breaking rates independently of each other 
as in the spin-blockade studies \cite{maisi2016, fujita2016, ptaszynski2017}, where the electron occupation preserving spin-flip rate was determined from the tunneling statistics. 

\begin{acknowledgments}
This work was performed as  part of the Academy of Finland Centre of Excellence program (project 312057). We acknowledge the provision of facilities by Aalto University at OtaNano - Micronova Nanofabrication Centre. V.F.M. thanks QuantERA project "2D hybrid materials as a platform for topological quantum computing" and NanoLund for financial support.
\end{acknowledgments}

\section{Supplemental Material}

The supplemental material contains details of the measurement setups used as well as additional data for Sample B.

\subsection{Measurement setup}

Sample A was measured in a homemade plastic dilution refrigerator with a base temperature of 50 mK (setup A), while sample B was measured in a dry dilution refrigerator with a base temperature of 20 mK (setup B). In both setups, the DC wires to the sample boxes were filtered with 1 m to 2 m of Thermocoax \cite{zorin1995}. Sample A was measured in a purely DC configuration with room-temperature voltage sources and resistive voltage dividers used to provide the gate and bias voltages shown in Fig. 2(a) of the main text. A room-temperature current amplifier with typically $10^9$ V/A gain was used to amplify the detector current, which was either recorded with a multimeter for basic characterization or sampled at 50 kHz for measuring charge detector time traces. 

\begin{figure}
\centering
\includegraphics{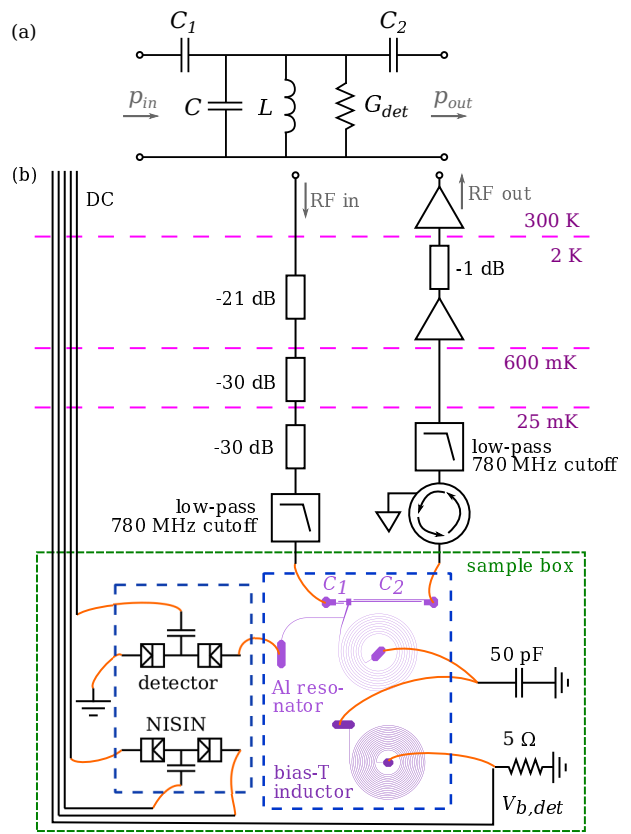}
\caption{(a) Equivalent circuit of the RF readout scheme used for sample B. (b) Details of the RF wiring in setup B.}
\label{rfsuppl}
\end{figure}

For the measurements in setup B, 
the DC gate voltages and bias voltages to the superconducting island are provided as in setup A, so that the characteristics of the superconducting island can be determined directly from I-V measurements. 
We use the charge detector as a radio-frequency SET \cite{schoelkopf1998}. The RF setup is an adaptation of that developed for radio-frequency normal-insulator-superconductor junction thermometers, described in Refs. \cite{gasparinetti2015, viisanen2015}. 
The detector SET is connected in parallel with a lumped-element $LC$ resonator with a resonance frequency of 600 MHz. We measure the transmitted power coupled through input and output capacitors $C_1$ and $C_2$, respectively, as shown in the schematic of Fig. \ref{rfsuppl}(a). The measured transmission at resonance depends on the differential conductance $G_{det}$ of the detector and thus the charge on the superconducting island. 
The input RF signal is supplied using heavily attenuated coaxial cables and filtered with a commercial low-pass filter at the mixing chamber, as shown in Fig. \ref{rfsuppl}(b). The output signal is amplified at 2 K and room temperature with a total gain of approximately 60 dB. We obtain traces of power versus time at a sampling rate of 100 kHz after analog and digital demodulation.
In the present setup, the sensitivity of the detector and thus the maximum charge detection bandwidth is limited by signal-to-noise ratio of the readout chain.
The resonator consists of a spiral inductor made out of 100 nm thick evaporated aluminum on a separate oxidized silicon chip and parasitic capacitance to ground. Another spiral inductor and a lumped-element capacitor on the sample stage allow applying a DC voltage across the SET,  while the other side of the SET is grounded at the sample stage. A 5 $\Omega$ resistor on the sample stage serves as a part of a cold voltage divider to provide the voltage bias for the detector. This biasing scheme prohibits measuring the current-voltage characteristics of the detector directly, but the detector current can be estimated from $V_{b,det}$ and the tunnel resistance at room temperature. We typically use DC bias voltages on the order of 400 $\mu$V and RF probe powers corresponding to an amplitude of 1 to 50 $\mu$V, depending on the DC bias point. Given that the RF probe amplitude and frequency are small compared to relevant energy scales, we expect the measurement setup not to affect the observed mechanism of detector backaction.

In all devices, a gold ground plane under the insulating aluminum oxide layer capacitively shunts the leads of the superconducting island. The ground plane acts as an on-chip filter to suppress photon-assisted tunneling caused by high-frequency microwave photons \cite{pekola2010}. The leads of the detector are left unshunted, so that the extra capacitance does not prohibit RF readout. The superconducting island itself is not on the ground plane, so that good capacitive coupling to the detector is possible. The capacitive coupling is provided by a 10 $\mu$m long chromium wire located under the aluminum oxide layer. The resistance of the wire is between 50 k$\Omega$ to 100 k$\Omega$, and the wire should therefore act as an $RC$ filter or resistive transmission line and filter out shot noise at high frequencies, while providing a good coupling at low frequencies. A similar wire was used for the coupling in Ref. \cite{saira2012}.

In setup A, the sample is mounted to the mixing chamber with a brass sample box with two nested caps, of which the inner cap is sealed with an indium seam and the outer is a threaded rotating cap. This setup has been previously used to reach the low quasiparticle densities reported in Refs. \cite{saira2012} and \cite{maisi2013}. In setup B, the sample is mounted in the mixing chamber in a copper box sealed with a single cap with an indium seal. We have additionally measured a third device with similar parameters in the same cryostat as sample B, but in a different sample holder with similar wiring as in setup A. This holder had an inner threaded cap and an outer indium-sealed cap, but with possibly leaky microwave connectors leading inside this outer cap. There, the measured quasiparticle tunneling rates corresponded to $\langle N_S \rangle \approx 3$. We conclude that a single rf-tight shield at the mixing chamber is both necessary and sufficient shielding for reaching the low quasiparticle densities presented here.

The digitized detector signal, either current or transmitted power, is digitally filtered with a fourth-order Butterworth low-pass filter with the cutoff frequency set to 5 kHz for the data shown for sample A and to 1 kHz for sample B.

\subsection{Occupation probabilities, tunneling rates and simulations for Sample B}

\begin{figure}[!h]
\centering
\includegraphics{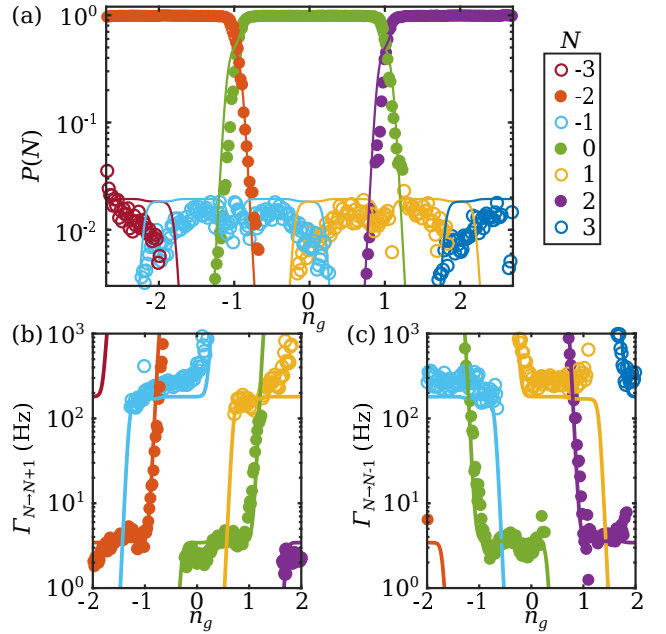}
\caption{Measured (circles) and simulated (solid lines) occupation probabilities (panel a) and transition rates between (panels b,c) charge states $N$ in Sample B at $V_{b,set}=0$, $V_{b,det} = 350$ $\mu$V and the probe power $p_{in} \approx$ -90 dBm. Simulation parameters are given in the text. Note that the vertical scale is logarithmic in all panels, in contrast to Fig. 3 of the main text.}
\label{analysisB}
\end{figure}

In Fig. \ref{analysisB} we show analysed data from Sample B, similar to that shown in Fig. 3 of the main text for sample A, at the same detector operating point as in the time traces of Fig. 2(g,h,i,j). The occupation probabilities of odd charge states are typically less than 3\%, with the tunneling rates out of the odd states typically around a few Hz. Solid lines are fits with parameters $R_T = 8.9$ M$\Omega$, $\Delta=210$ $\mu$eV and $E_C = 95$ $\mu$eV extracted from large-scale current-voltage characteristics and the pair-breaking rate $\Gamma_{pb}=100$ Hz as the only free parameter. Since the recombination rate $\Gamma_{rec} = 9.7 $ kHz is much larger than the measured tunneling rates in both devices, the main effect of a reduced $\Gamma_{pb}$ in the simulations is to decrease the quasiparticle population and consequently the tunneling rates, which depend linearly on the quasiparticle population. Compared to the data shown in Fig. 3 of the main text, $\Gamma_{pb}$ is reduced by a factor of 50,  and the measured tunneling rates $\Gamma_{qp}(\langle N_{S,even} \rangle)$ out of even charge states are smaller by a similar factor when taking into account the different $R_T$ of the two devices.
We note that two-electron Andreev tunneling events are not included in the model. Especially around $n_g = N$ with $N$ even, Andreev tunneling is expected to influence the occupation probabilities more than in sample A because the single-electron tunneling rates are smaller in this device, while the Andreev rates are larger than in sample A due to a smaller $R_T$. 

\end{document}